\documentclass[12pt,twoside]{article}
\usepackage{amssymb}
\def\R{{\mathbb R}}
\def\N{{\mathbb N}}

\def\supp{{\, \mbox{\rm supp}\,}}
\begin{document}
\baselineskip=22pt
\pagestyle{myheadings} \markboth{ S. Albeverio, H. Gottschalk and
M. W. Yoshida }{Representing EQFT via particle systems} \thispagestyle{empty}

\title{Representing Euclidean quantum fields as scaling limit of particle systems}

\author{Sergio Albeverio and Hanno Gottschalk\footnote{Supported by D.F.G. Project 'Stochastic analysis
 and systems with infinitely many degrees of freedom'}\\
{\small Institut f\"ur angewandte Mathematik,} \\ {\small Wegelerstr. 6, D-53115
Bonn, Germany}\\{\small e-mail: albeverio@uni-bonn.de /
gottscha@wiener.iam.uni-bonn.de}\\
\\
Minoru W. Yoshida\footnote{Partially supported by Grant-in-Aid Research No. 12640159 Japanese Ministry of Education and Science.}\\
{\small Department of Mathematics and Systems Engeneering,}\\
{\small The University of Electrocommunications 1-5-1,}\\
{\small Chofugaoka, Tokyo 182-8585, Japan}\\
{\small e-mail: yoshida@se.cas.uec.ac.jp}}
\maketitle
{\noindent \small {\bf Abstract.} We give a new representation of Euclidean quantum fields as scaling
limits of systems of interacting, continuous, classical particles in the grand canonical ensemble.
}

 \section{Introduction}
The success of the Euclidean strategy in constructive quantum field theory (QFT) in the 1970s
in many aspects was due to the formal analogy of Euclidean QFT with classical statistical mechanics, cf. \cite{Si}
and references therein. This analogy has been used in technically different approaches, as e.g. the lattice approximation \cite{Si},
the identification of the (massive) sine-Gordon model with the Coulomb (Yukawa) gas \cite{AHK1,F} or, more
recently, the polymer resp. random walk representation \cite{AFHKL} resp. \cite{FFS}.
In the present letter we introduce a new and conceptually different representation of Euclidean quantum fields by the means of systems of classical continuous
particles in the grand canonical ensemble (GCE).

In \cite{AGW1} it was realized that a 'Poissonization' of the natural stochastic partial pseudo-differential equation associated to the Euclidean
Bosonic free field \cite{N1} leads to a new, in general not reflection
positive, Euclidean QFT which however permits an analytic continuation (in the sense of quantum fields with indefinite metric \cite{MS}) from imaginary Euclidean time to real,
relativistic time. Here we show that these 'Poisson' Euclidean quantum fields can be identified with a noninteracting particle system in the grand canonical ensemble
(Section 2).

Furthermore, using that the path-properties of 'Poisson' Euclidean QFTs are more regular than in the standard 'Gaussian' case, we introduce a class of
ultra-violet (UV) finite, local interactions (in arbitrary dimension) for these models. On the level of classical
particles moving in continuous space-time and described in the GCE, such field-theoretic interactions can be
interpreted as a potential energy in the classical Hamiltonian of the $n$-particle system (Section 3).

Finally we identify a scaling limit under which the interacting 'Poisson' Euclidean QFTs formally converge to the related perturbed Gaussian models. On the level of
particle systems this corresponds to a scaling of activity and charge, while on the level of quantum fields there is a strong formal analogy with the block-spin formulation
of the renormalization group, see e.g. \cite{FFS}. We identify the change of path-properties under this limit and we give arguments to show, how in some simple situations this can be seen
as a source of triviality, from which the necessity of a nontrivial renormalization arises (Section 4).

\section{Convoluted Poisson noise}
Let $d\geq 2$ be the space-time dimension. The Euclidean (neutral, scalar, Bosonic) free field $\phi_0^g(x)$, $x\in\R^d$, \cite{N1} of mass $m>0$ in $d$ dimensions\footnote{The subscript
zero in this letter stands for noninteracting fields, not for time zero fields.} can be obtained from the stochastic pseudo differential equation
\begin{equation}
\label{2.1eqa}
(-\Delta+m^2)^{1/2}\phi^g_0(x)=\eta^g_0(x)
\end{equation}
where $\eta^g_0$ is a Gaussian white noise of intensity $\sigma>0$, i.e. a centered Gaussian random field with covariance function $\sigma^2\, \delta(x-y)$. Let $G(x)$ be the Green's
function associated to $(-\Delta+m^2)^{1/2}$, then $\phi^g_0(x)=G*\eta^g_0(x)$ defines a pathwise solution of (\ref{2.1eqa}), i.e. a solution for any fixed random parameter. The convolution
$G*\eta_0^g(x)$ is well-defined, since by Minlos' theorem \cite{M} $\eta_0^g(x)$ is a (random) tempered distribution and $(-\Delta+m^2)^{1/2}$ is continuously invertible on the space of tempered
distributions.

Equation (\ref{2.1eqa}) can be modified by replacing the noise $\eta_0^g$ by a Poisson noise as follows: For a bounded region $\Lambda\subseteq \R^d$ we define
the Poisson noise random field on $\Lambda$ as
\begin{equation}
\label{2.2eqa}
\eta_{0,\Lambda}(x)=\sum_{j=1}^{N_\Lambda^z}S_{\Lambda,j}\, \delta(x-Y_{\Lambda,j})
\end{equation}
where $N_\Lambda^z$ is a Poisson random variable with intensity $z|\Lambda|$, i.e. $P\{N^z_\Lambda=n\}=e^{-z|\Lambda|}(z|\Lambda|)^n/n!$. The parameter $z>0$ is called the
activity of the noise. $\{S_{\Lambda,j}\}_{j\in\N}$ is a family of independently identically distributed (i.i.d.) real-valued random variables with distribution given by a probability
measure $r$ on $\R$. For simplicity we assume that $\supp r\subseteq[-c,c]$ for some $c>0$. Finally, $\{ Y_{\Lambda,j}\}_{j\in\N}$ is a family
of $\R^d$-valued i.i.d. random variables distributed according to the uniform distribution on $\Lambda$.

We chose $\Lambda_n\subseteq \R^d$ a sequence of bounded, disjoint subsets of $\R^d$ such that $\cup_{n=1}^\infty\Lambda_n=\R^d$. We then define
\begin{equation}
\label{2.3eqa}
\eta_0(x)=\sum_{n=1}^\infty\eta_{0,\Lambda_n}(x)
\end{equation}
where the noises $\eta_{0,\Lambda_n}(x)$ are independent from each other for different $n$. One can verify\footnote{One way to see this is to calculate the Fourier transform of $\langle\eta_0,f\rangle$, $f$ a Schwartz test function, and show 
that it does not depend on $\Lambda_n$. The statement then follows from the uniqueness result of Minlos' theorem \cite{M}, cf. Section 3.2 of \cite{AGY} for the details. } 
that the so-defined Poisson noise $\eta_0(x)$ does not depend on the choice of
the sets $\Lambda_n$ and that the restriction of $\eta_0(x)$ to a bounded set $\Lambda\subseteq \R^d$ has a representation (\ref{2.2eqa}).  Furthermore, $\langle\eta_0,f\rangle=\int_{\R^d}\eta_0(x)f(x)\, dx$
for any Schwartz test function $f$ exists with probability one, since $\eta_0(x)$ with probability one is polynomially bounded, as a consequence of the Borel-Cantelli lemma.
$\langle\eta_0,f\rangle$ has infinitely divisible (probability) law and $\langle\eta_0,f\rangle$ and $\langle\eta_0,h\rangle$ are independent from each other whenever the test functions
$f$ and $h$ have disjoint support. Also, $\eta_0(x)$ is invariant under Euclidean transformations (translations, reflections and rotations) in law.

 These properties, which trivially also hold for the Gaussian white noise $\eta_0^g(x)$, show that from a mathematical point of view a legitimate modification of Eq. (\ref{2.1eqa}) is obtained
if we replace the Gaussian noise $\eta_0^g(x)$ in (\ref{2.1eqa}) with $\eta_0(x)$. The solution $\phi_0(x)=G*\eta_0(x)$, constructed as above and called convoluted Poisson noise (CPN), is a 
'Poissonian' modification of the Euclidean free field.

This mathematical argument deserves a physical justification: As proven in \cite{AGW1}, the Euclidean quantum fields defined as CPN can be analytically continued to relativistic quantum fields
with indefinite metric \cite{MS}. Given that the covariance of CPN coincides with the covariance of the Euclidean free field\footnote{With intensity $\sigma^2=z^2\int_{\R}s^2\, dr(s)$. This should not be mixed up
 with the relations for the scaling limit in Section 4.}, the asymptotic particle structure of these quantum fields \cite{AG} consists of free, relativistic particles of mass $m$
and the scattering behavior is trivial, since mass-shell singularities of the higher order truncated Wightman functions are infra-particle like \cite{Schr}. In this sense CPN is still a scattering free field. These truncated Wightman functions can however be decomposed into
direct integrals (over the mass-parameter) of Wightman functions with non-trivial scattering, \cite{AG,AGW1}. Further motivation in favor of 'Poissonization' will be given in Section 4.

To finish this section, we consider the path properties of the convoluted Poisson noise
 field $\phi_0(x)$: From the exponential decay $|G(x)|\sim e^{-m|x|}$ as $x\to\infty$ and the Borel-Cantelli Lemma
 one gets that with probability one the expression
\begin{equation}
\label{2.4eqa}
\phi_0(x)=\sum_{n=1}^\infty\sum_{j=1}^{N_{\Lambda_n}^z}S_{j,n}\, G(x-Y_{\Lambda_n,j})
\end{equation}
is summable if $x$ does not coincide with any of the $Y_{\Lambda_n,j}$. Only in the latter points the random field $\phi_0(x)$ has locally integrable singularities
$\sim |x-Y_{\Lambda_n,j}|^{-d+1}$ as $G(x)\sim |x|^{-d+1}$ at $0$. By this analysis we conclude that $\phi_0(x)$ with probability one is given by
a locally integrable function. This is the basic difference with the Gaussian case, where
the path properties of the Euclidean free field depend crucially on the dimension and for $d\geq 2$ cannot be represented by functions, cf. e.g. \cite{RR}.

\section{Local interactions and interpretation as particle systems}
Let $v:\R\to\R$ be a function such that $|v'(\phi)|<C$ $\forall \phi\in\R$. Then $v$ induces a (nonlinear) transform on the space of
locally integrable functions\footnote{Symbols as $\phi$ and $\eta$ without subscripts and superscripts are used for sample paths (hence non-random objects) in the space of locally integrable functions and
signed measures with finite support (in $\Lambda$), respectively, and are used as integration variables in (rigorous) integrals over such path-, respectively charge-, configurations.
The random processes \cite{M} related to path- or configuration space measures (defining these integrals) are denoted by $\phi_0,\phi_{0,\Lambda}$ etc., resp. $\eta_0,\eta_{0,\Lambda}$, depending on the measure.
Likewise, $s_j,y_j,n$ etc. are non random coordinates parametrizing $\eta$ while $S_{j,\Lambda},Y_{j,\Lambda},N_\Lambda^z$ are the related random objects parametrizing $\eta_{0,\Lambda}$}
via $\phi(x)\to v(\phi(x))$. Let now $\nu_0$ be the probability measure on the space of locally integrable functions
associated with the convoluted Poisson white noise $\phi_0(x)$ \cite{AGW1,M}. Then
\begin{equation}
\label{3.1eqa}
d\nu_\Lambda(\phi) ={\exp\{ -\lambda\int_\Lambda v(\phi(x))\, dx\}\over\int \exp\{ -\lambda\int_\Lambda v(\phi(x))\, dx\}
\, d\nu_0(\phi)}\,d\nu_0(\phi)
\end{equation}
defines a new measure $\nu_\Lambda$ on the space of locally integrable functions. Here, $\Lambda\subseteq\R^d$ is a bounded region and plays the role of an infra-red cut-off and $\lambda\geq 0$
is the coupling constant. As $v(\phi(x))$ is locally integrable, the expression (\ref{3.1eqa}) requires no further regularization, not even Wick ordering. Thus, for convoluted Poisson
noise there exists a class of local potentials which is UV-finite independently of the dimension $d\geq 1$. This is in striking contrast to the Gaussian case. It should however be mentioned
that the condition $|v'|<C$, which also implies that the denominator in (\ref{3.1eqa}) is finite, excludes polynomial potentials requiring  regularization also in the Poisson case. However, trigonometric potentials,
as the sine-Gordon potential, used in \cite{AHK1,F} as perturbations of free fields, exist in our Poisson case independently of the dimension.

We now want to give a statistical mechanics interpretation to (\ref{3.1eqa}): $\eta_{0,\Lambda}(x)$ in (\ref{2.2eqa}) is interpreted as
the random process describing $N_\Lambda^z$ classical, indistinguishable and noninteracting
particles with positions $Y_{\Lambda,j}$ in the 'box' $\Lambda$. The particle with the position $Y_{\Lambda,j}$ carries the charge
$S_{\Lambda,j}$. As $N_\Lambda^z$ is Poisson distributed, the particle system is in the (configurational) grand canonical ensemble (GCE)
with activity $z$. The Euclidean Poisson quantum field $\phi_{0,\Lambda}(x)=G*\eta_{0,\Lambda}(x)$ has a natural interpretation as
the static field of the charge configuration $\eta_{0,\Lambda}(x)$, i.e. a unit charge in the point $y$ gives rise to a static field $G(x-y)$
and the static field of a number of charges is then obtained by superposition.

Next we extend this statistical mechanics interpretation to the interacting case. Obviously, the interaction in (\ref{3.1eqa}) is given by the nonlinear
energy density $v$ of the static field. To formulate this on the level of the associated particle system, we find it more
convenient to consider the particle system in the box $\eta_{0,\Lambda}(x)$ and the associated static field $\phi_{0,\Lambda}(x)$ instead
of using the infra-red cut-off in (\ref{3.1eqa}). Identifying $\eta(x)=\sum_{j=1}^ns_j\delta(x-y_j)$ with the non-ordered $n$-tuple
$\{ (y_1,s_1),\ldots,(y_n,s_n)\}$ we obtain as the interaction term in the $n$-particle Hamiltonian (for the normalization $v(0)=0$)
\begin{eqnarray}
\label{3.3eqa}
U(\eta)&=&\int_{\R^d}v(G*\eta(x))\, dx\nonumber\\
&=&U(y_1,\ldots,y_n;s_1,\ldots,s_n)\nonumber\\
&=&\int_{\R^d}v\left(\sum_{j=1}^ns_jG(x-y_j)\right)dx\, ,
\end{eqnarray}
where the integral exists due to the exponential decay of $G(x)$ as $|x|\to\infty$.

Let $\nu_{0,\Lambda}$ be the measure associated with the Euclidean quantum field $\phi_{0,\Lambda}(x)$ and let $\mu_{0,\Lambda}$ be
the measure associated with the particle system in the box $\eta_{0,\Lambda}(x)$. We then define in analogy with (\ref{3.1eqa}), however, as we are in a finite volume,
without additional infra-red cut-off
\begin{equation}
\label{3.4eqa}
d\tilde \nu_\Lambda(\phi) =
{\exp\{ -\lambda\int_{\R^d} v(\phi(x))\, dx\}\over\int \!\exp\{ -\lambda\int_{\R^d} v(\phi(x))\, dx\}\,
d\nu_{0,\Lambda}(\phi)}\, d\nu_{0,\Lambda}(\phi)
\end{equation}
and likewise we define the Gibbs measure in the finite volume $\Lambda$ as
\begin{equation}
\label{3.5eqa}
d\tilde \mu_\Lambda (\eta)={\exp\{-\lambda U(\eta)\}\over \int \exp\{-\lambda U(\eta)\}\,d\mu_{0,\Lambda}(\eta)}\, d\mu_{0,\Lambda}(\eta)
\end{equation}
where $\eta(x)=\sum_{j=1}^ns_j\delta(x-y_j)$ with $n\in\N_0$, $y_j\in\Lambda$ and $s_j$ in the support of the measure $r$. Let $\tilde\phi_\Lambda(x)$ be the Euclidean quantum field
associated with the measure $\tilde \nu_\Lambda$ and $\tilde\eta_\Lambda(x)$ be the interacting particle system associated with $\tilde  \mu_\Lambda$. Then, using (\ref{3.3eqa})--(\ref{3.5eqa})
one obtains by setting $\phi=G*\eta$ that $\tilde\phi_\Lambda(x)$ and $\tilde\eta_\Lambda(x)$ are related via the generalized Poisson equation (\ref{2.1eqa}), hence $\tilde \phi_\Lambda(x)$
is given by the static field of the interacting particle system $\tilde \eta_\Lambda(x)$.

One can now use techniques from the theory of Gibbs measures of classical continuous particle systems to study the
infinite volume limit $\Lambda\uparrow\R^d$, cf. \cite{Ru} for the general strategy and \cite{AGY} for some first steps in that direction.

Interestingly, one can formulate a number of well-known systems of statistical mechanics as interacting Poisson quantum fields. Let $r=\delta_1$, i.e.
all charges take the value $+1$. Then e.g. the choice $G=\chi_{B_R}$, the characteristic function of the ball centered at zero with radius $R$, and
$v(\phi)=0$ if $\phi<2$, $v(\phi)=\infty$ if $\phi\geq 2$ gives a quantum field theoretic definition of the gas of hard spheres, cf. (\ref{3.3eqa}). Likewise, taking $G$ an arbitrary smooth function of exponential decay
with a singularity at $0$ and $v(\phi)=0$ if $\phi<C$, $C>0$, and $v(\phi)=1$ if $\phi\geq C$, respectively  $v(\phi(x))=\delta(\phi(x)-C)|\nabla \phi(x)|$, we obtain the field theoretic
potentials defining the 'trigger potential', 'isodensity contour potential' respectively, from stochastic geometry \cite{SKM}. Lastly, smearing out $G$ with a mollifier, we can see that the quadratic potenial
$v(\phi)=\phi^2$ up to a chemical potential yields a particle potential with two point interactions\footnote{In fact, a polynomial potential of degree $n$ gives a potential which includes only $j$-point interactions for $j\leq n$.
However, renormalization of such potentials for $n>2$, when removing the mollifier, is more delicate.} . Removing the mollifier, the chemical potential diverges and has to be renormalized similarly as in classical electrostatics.
Further details on the connections with classical statistical mechanics can be found in \cite{AGY}.

\section{The Gaussian scaling limit}
In this section we consider the scaling limit $z\to\infty$ with field strength renormalization $\phi\to\phi/\sqrt z$. On the level of particle systems this can be seen as the scaling $z\to\infty$, i.e. the average number of particles per
volume goes to infinity, and the charges of the particles scale with $1/\sqrt z$. We indicate this scaling with a superscript $z$.

If the particle system is noninteracting and the gas of particles is neutral
 in average, i.e. $\int_{\R}s\,dr(s)=0$, then, by the central limit theorem, we get for a Schwartz test function $f$
  (note that $N_\Lambda^z$ has expectation $z|\Lambda|$ and $S_{\Lambda,j}f(Y_{\Lambda,j})$ are independent, identically distributed random variables with expectation zero and variance
  $\int_{\R}s^2\, dr(s)\,\int_\Lambda f^2(x)\, dx/|\Lambda|^2$),
\begin{equation}
\label{4.1eqa}
\langle\eta_{0,\Lambda}^z,f\rangle={1\over\sqrt z}\sum_{j=1}^{N_\Lambda^z} S_{\Lambda,j}f(Y_{\Lambda,j})
\stackrel{\cal L}{\to} \langle \eta_{0,\Lambda}^g,f\rangle
\end{equation}
where $\eta_{0,\Lambda}^g(x)=\chi_\Lambda(x)\eta_0^g(x)$, $\sigma^2=\int_{\R}s^2\, dr(s)$
and $\stackrel{\cal L}{\to}$ stands for convergence in probability law.

Likewise, using also (\ref{2.3eqa}) and (\ref{2.4eqa}), we get that $\langle\eta_0^z,f\rangle\stackrel{\cal L}{\to}\langle\eta^g_0,f\rangle$ and
$\langle\phi^z_0,f\rangle\stackrel{\cal L}{\to}\langle\phi_0^g,f\rangle$. The latter scaling can  equivalently be understood as a scale and mass transformation $x\to\alpha x$, $m\to m/\alpha$ in
conjunction with a field strength renormalization $\phi\to \alpha^{(d-2)/2} \phi$, with $\alpha=z^{1/d}\to\infty$, cf. \cite{AGY}. This clearly is analogous to the block-spin transformation in lattice Euclidean QFT \cite{FFS}. In the
general sense that the renormalization group describes a process adding more and more micro-structures to a finite volume,
we can thus consider the scaling $z\to\infty$ as an adequate implementation of the renormalization group in our framework.

Next, we want to consider the scaling for interacting models starting with UV-regularized models where $G$ is replaced by $G_\epsilon=G*\chi_\epsilon$ with
a mollifier $\chi_\epsilon\to\delta$ as $\epsilon\to+0$. We can construct a measure $\nu_\Lambda^{z,\epsilon}$ as in (\ref{3.1eqa}) where $G$ is replaced by $G_\epsilon$. As the theory now is UV-regularized, the related perturbed Gaussian measure $\nu_\Lambda^{g,\epsilon}$ exists. For simplicity we also assume that $v$ is a bounded function.
Let $\phi_\Lambda^{z,\epsilon}(x)$ and $\phi_\Lambda^{g,\epsilon}(x)$ be the Euclidean quantum fields (random fields) determined by the measures $\nu^{z,\epsilon}_\Lambda$ resp. $\nu^{g,\epsilon}_\Lambda$. We want to show that $\langle \phi_\Lambda^{z,\epsilon},f\rangle\stackrel{\cal L}{\to}\langle\phi_\Lambda^{g,\epsilon},f\rangle$ as $z\to\infty$, i.e. that
the UV-regularized perturbed Poisson models converge to the related perturbed Gaussian ones. To see this, we consider the Fourier transform of $\langle\phi_\Lambda^{z,\epsilon},f\rangle$, with $t\in\R$
\begin{equation}
\label{4.2eqa}
{\int \exp\{it\langle\phi,f\rangle-\lambda\int_\Lambda v(\phi(x))dx\}\,d\nu_{0}^{z,\epsilon}(\phi)\over\int \exp\{-\lambda\int_\Lambda v(\phi(x))dx\}\,d\nu_{0}^{z,\epsilon}(\phi)}
\end{equation}
Expanding the numerator in increasing powers of $\lambda$ we get as the $n$-th coefficient 
\begin{eqnarray}
\label{4.3eqa}
&&{1\over n!}\int_{\Lambda^{\times n}}\int \exp\{it\langle\phi,f\rangle\} v(\langle\phi,\chi_{\epsilon,y_1}\rangle)\cdots\nonumber\\
&&\hspace{2cm}\times \cdots \,v(\langle\phi,\chi_{\epsilon,y_n}\rangle)\, d\nu_0^z(\phi)\, dy_1\cdots dy_n
\end{eqnarray}
where $\chi_{\epsilon,y}(x)=\chi_\epsilon(x-y)$ and $\nu_0^z$ is the measure associated to $\phi_0^z(x)$ (where the latter random field has no UV-cut-off). As
$\phi_0^z(x)$ converges in law to $\phi^g_0(x)$ as $z\to\infty$, we can replace in (\ref{4.3eqa}) $d\nu^z_0(\phi)$ with $d\nu^g_0(\phi)$ in that limit. Furthermore,
as the expansion in $\lambda$ converges uniformly (as $v$ is bounded) and the the denominator has a corresponding expansion with $f=0$, we can
replace in (\ref{4.2eqa}) $d\nu^{z,\epsilon}_0(\phi)$ with $d\nu_0^{g,\epsilon}(\phi)$ as $z\to\infty$, which establishes our claim.

We now want to consider the same situation, but without UV-cut-off. Furthermore, we concentrate ourselves on trigonometric potentials \cite{AHK1}
\begin{equation}
\label{4.4eqa}
v(\phi)=\int_{\R}\cos(\alpha\phi)\, d\rho(\alpha)
\end{equation}
with $\rho$ a finite signed measure on $\R$ with $\rho\{0\}=0$. One can then show that $\langle\phi_\Lambda^{z},f\rangle\stackrel{\cal L}{\to}\langle\phi_0^g,f\rangle$, i.e.
the scaling limit of the perturbed Poissonian model is Gaussian (and thus physically trivial)\cite{AGY}. Here we want to give a qualitative argument to explain this statement:
The average path $\phi(x)$ in the support of $\nu_0^z$ with the increasing numbers of particles per volume of the associated particle system has more and more points where the path
is behaving like $G(x)/\sqrt z$ for $x\to 0$. As $G$ at $0$ is not square integrable, these singularities have enough 'volume' to make the average path more and more singular, as
due to the process $z\to\infty$ more and more particles with charge $\sim\pm 1/\sqrt z$ are being added. Finally, all the area of $\Lambda$ becomes covered with singularities and the average absolute
value of
\begin{equation}
\label{4.5eqa}
\int_\Lambda v(\phi(x))\,dx=\int_{\R}\int_\Lambda \cos(\alpha \phi(x))\, dx \, d\rho(\alpha)
\end{equation}
becomes smaller and smaller as the oscillations in the inner integral integrate out to zero. In the limit $z\to\infty$ then the potential converges to zero (in law). This process eludicates the need of
a renormalization procedure for the potential in the limit $z\to\infty$, i.e. in order to avoid triviality one has to consider energy densities $v_z$ depending on the scaling parameter $z$.

The easiest case of such a renormalization is the coupling constant renormalization of the sine-Gordon model in $d=2$ dimensions, see e.g. \cite{F}. There, $v(\phi)=\cos(\alpha \phi)$ for $\alpha$ sufficiently small. To renormalize
this potential (which otherwise would lead to a trivial scaling limit) one sets
\begin{equation}
\label{4.6eqa}
v_z(\phi)=:\cos(\alpha \phi):^z={\cos(\alpha\phi)\over\int \cos(\alpha\varphi(0))\, d\nu_0^z(\varphi)}\,.
\end{equation}
The $:~:^z$-term in the middle is defined by the right hand side in the spirit of the Wick-ordering of the cosinus function in the Gaussian case \cite{AHK1,F}.
 
One can then show that in a certain expansion the characteristic functional (see (\ref{4.2eqa})) of the process $\phi_\Lambda^z(x)$ with interaction given by $v_z$ converges in any order of that expansion
to the related expression for the sine-Gordon model, cf. \cite{AGY} for further details.

\vspace{.2cm}
\noindent {\bf Acknowledgements.} We thank Klaus R. Mecke and Tobias Kuna for interesting discussions.


\begin{thebibliography}{11}
\bibitem{AFHKL} S. Albeverio, J. E. Fenstad, R. H\o egh-Krohn, T. Lindstr\o m: Nonstandard Methods inStochsatic analysis and Mathematical Physics, Academic Press 1986 (Russian Transl. Mir 1988).
\bibitem{AG} S. Albeverio, H. Gottschalk: Commun. Math. Phys. {\bf 261}, 491--513 (2001).
\bibitem{AGW1} S. Albeverio, H. Gottschalk, J.-L. Wu, Rev. Math Phys., Vol {\bf 8}, No. {\bf 6},  763--817, (1996), Commun. Math. Phys. {\bf 184},  509--531, (1997).
\bibitem{AGY} S. Albeverio, H. Gottschalk, M. W. Yoshida: Systems of classical particles in the grand canonical ensemble, scaling limits and quantum field theory. Bonn 2001, \underline{www.lqp.uni-goettingen.de/papers/01/05}.
\bibitem{AHK1} S. Albeverio, R. H\o egh-Krohn, Commun. Math. Phys. {\bf 30}, 171--200 (1973).
\bibitem{FFS} R. Fern\`andez, J. Fr\"ohlich, A. D. Sokal:
Random Walks, Critical Phenomena and triviality in Quantum Field Theory,
Berlin / Heidelberg / New York: Springer-Verlag 1992.
\bibitem{F} J. Fr\"ohlich, Commun. Math. Phys. {\bf 47}, 233--268 (1976).
\bibitem{M} R. A. Minlos, Transl.
Math. Stat. and Probability , AMS Providence,{\bf 3}, p. 291--313 (1963).
\bibitem{MS} G. Morchio, F. Strocchi Ann.
Inst. H. Poincar\'e, {\bf 33}, 251--282 (1980).
\bibitem{N1} E. Nelson,
J. Funct. Anal. {\bf 12}, 97--112 (1973),
J. Funct. Anal. {\bf 12}, 211--227 (1973)
\bibitem{RR} M. Reed, J. Rosen, Commun. Math. Phys. {\bf 36}
123--132 (1974).
\bibitem{Ru} D. Ruelle: Statistical mechanics -- rigorous results. Benjamin, London / Amsterdam / Don Mills (Ontario) /
Sydney / Tokyo 1969.
\bibitem{Schr} B. Schroer, Fortschr. Phys. {\bf 1/63} 1--31 (1963)
\bibitem{Si} B. Simon: The $P(\phi )_2$ Euclidean (Quantum) Field Theory. Princeton University Press, Princeton, New Jersy, 1974.
\bibitem{SKM} D. Stoyan, W. S. Kendall, J. Mecke: Stochastic geometry and its applications. Wiley \& Sons, 1987.
\end{thebibliography}
\end{document}